\documentclass[10pt]{iopart}
\usepackage{epsfig}  
\usepackage{wrapfig}

\begin{document}

\title[New results from CERES]{New results from CERES}

\author{A.~Mar\'{\i}n for the CERES
Collaboration \footnote[1]{For the full CERES Collaboration author list and
acknowledgments, see Appendix "Collaborations" of this volume.} }

\begin{abstract}

During the year 2000 the CERES experiment, upgraded with a radial drift
TPC, took a large data sample of Pb on Au
collisions at 158~AGeV triggered on the ~8\% most central collisions. 
A very detailed calibration of the radial drift TPC was since completed.
Preliminary results on $e^+e^-$ pairs and $\phi$ mesons reconstructed in
the $K^+K^-$ channel are presented.

\end{abstract}

\section{Introduction}

QCD calculations predict a 
transition from ordinary hadronic matter into a
plasma of deconfined quarks and gluons at high energy density. Dileptons, which have 
negligible final state interactions, represent a very
suitable probe for the study of this new state of matter. 
The CERES/NA45 experiment at the CERN SPS has made major contributions to the measurement of low-mass 
e$^+$e$^-$ pairs in ultra-relativistic heavy-ion collisions. 
CERES has measured an enhanced dilepton production in the invariant mass region 
$m_{e^+e^-}>$~0.2~GeV/c$^2$ in S+Au at 200~AGeV \cite{exp} and 
in Pb+Au at 158~AGeV \cite{nantes} and at 40~AGeV
\cite{40gev} 
compared to the contribution from known hadronic sources. The enhancement is 
absent in p-induced reactions \cite{pp}. Pion annihilation has been taken into
account as an additional mechanism for e$^+$e$^-$ production but the 
experimental spectra cannot be explained without introducing medium modifications of 
vector mesons, particularly of the $\rho$. The question of possible
modifications of other vector mesons, specially of the $\omega$ and the
$\phi$, and the role of chiral symmetry restoration remained however open.

In order to further investigate the enhancement, the CERES spectrometer
was upgraded during 1998 by the addition of a Time Projection
Chamber (TPC) with radial electric drift field \cite{up2,up1,qm99} which 
improves the mass resolution and the electron identification. 
During the year 2000 CERES took a large data sample consisting of
30$\cdot10^6$ and 3$\cdot10^6$ events of Pb on Au collisions at 158 AGeV triggered on the
8\% and 20\% most central collisions, respectively. 
The challenge was to calibrate the highly complex radial drift
TPC to bring the momentum resolution to the design limit.

\section{Experimental Setup}
\begin{figure}
\hspace*{1cm}\mbox{\epsfig{file=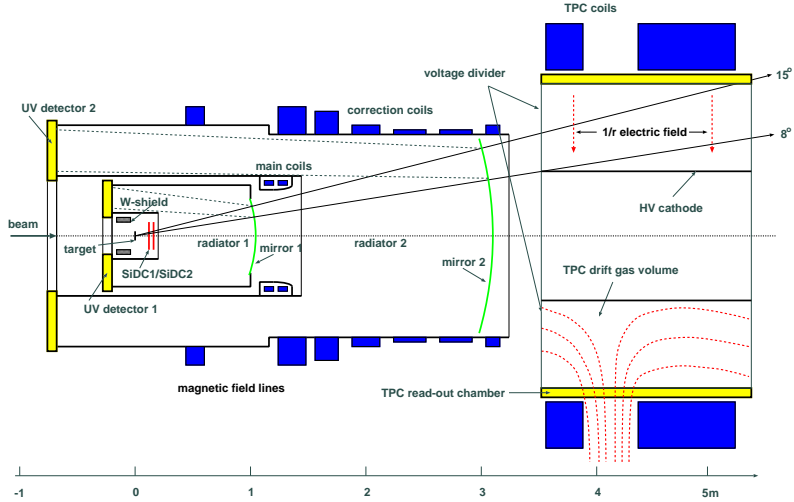,angle=-90,width=0.8\textwidth}}
\caption{Cross section through the upgraded setup of the CERES
spectrometer}
\label{fig:setup}
\end{figure}

The CERES experiment (Fig.~\ref{fig:setup}) is optimized to measure low mass electron pairs close to 
mid-rapidity (2.1$<\eta<$2.6) with full azimuthal coverage. 
A vertex telescope, composed of two Silicon Drift Detectors (SDD) positioned
at 10 cm and 13.8 cm downstream of a segmented Au target, provides a precise vertex reconstruction, 
angle measurement for charged particles and rejection of close pairs
from $\gamma$ conversions and $\pi^0$ Dalitz decays. Two Ring Imaging CHerenkov (RICH) detectors, 
operated at a high threshold ($\gamma_{th}$=32), are used for electron 
identification in a large hadronic background. The new radial-drift TPC,
positioned downstream of the original spectrometer, has an active length of 2 m 
and a diameter of 2.6~m. A gas mixture of  Ne (80\%) and CO$_2$ (20\%)
is used. It is operated inside a magnetic field
($\vec{B}$, indicated by the dashed field lines inside the TPC in Fig.~\ref{fig:setup}) 
with a maximal radial component of 0.5 T and provides up to 20
space points for each charged particle track. This 
is sufficient for the momentum determination and for additional
electron identification via d$E$/d$x$ in the TPC.
In the configuration with the TPC, the magnetic field between the
two RICH detectors is switched off and there is thus no deflection
between them, allowing to use them in a combined mode resulting in an increased 
electron efficiency. 
Moreover, the TPC opens new possibilities to study hadronic 
observables~\cite{qm01,hbt,hbt1,fluc,fluc1}.

\section{Calibration of the TPC}
One of the main tasks of the TPC calibration was to understand the
electric field ($\vec{E}$) which is dominantly radial with
$E_r$ $\sim$ $1/r$ but has a small azimuthal component due to the polygonal shape of 
the TPC and small longitudinal components at the end caps. 
A custom program based on the relaxation method was developed 
for calculating the electric field including 
the proper knowledge of the field cage resistors, the field
distortions caused by 
slightly displaced 
chambers and the leakage of the amplification field through the cathode
wire plane.

The magnetic field in the TPC is generated by two coils running with opposite
currents (Fig.\ref{fig:setup}). The two main components $B_z$ and $B_r$ 
change with $r$ and $z$. A field map was measured before the 
installation of the TPC in the experimental area. 
Compared to the nominal magnetic field calculated with the POISSON
program the measured field deviates from azimuthal symmetry by a few \%. 
These deviations are now included as corrections to the nominal field map.

The electron mobility as a function of $\vec{E}$ 
was determined from laser events containing tracks at known positions
for a given set of gas parameters.
Relative run-by-run variations caused by changes in composition,
temperature, and pressure (monitored by the slow
control system of the experiment) are calculated with the MAGBOLTZ program \cite{gar}.
The transformation from pad, time, and plane to the laboratory
coordinate system ($x$, $y$, and $z$) is done using a fourth-order 
Runge-Kutta method. The drift trajectory is calculated 
starting at the cathode plane using in each point the drift velocity 
vector, $\vec{v}_D = \mu { {1} \over{1+(\mu |B|)^2 } } 
( \vec{E} + \mu ( \vec{E} \times \vec{B} ) + \mu^2 ( \vec{E} \cdot
\vec{B} ) \vec{B} )$ where the components parallel to the electric field and parallel to
($\vec{E} \times \vec{B}$) have been modified to account for the
difference observed between using  $\vec{v}_D$ from the above equation
and the Monte Carlo drift
option in the MAGBOLTZ program \cite{gar}.
To fully account for the Lorentz angle a correction is
extracted from the symmetry of the 1/p distribution of
identified pions under the assumption that very high momentum particles
have vanishing curvature.
After all these corrections the local distortions are small compared
to the resolution.

The momentum of tracks from the target is determined from the combination
(resolution-weighted) of a 3-parameter fit that takes into 
account multiple scattering before the TPC (dominant for low-momentum particles)
and a 2-parameter fit that assumes that tracks come from the vertex
(dominant for high-momentum particles). The momentum resolution
depends on the number of hits on a track
and on the single-hit position resolution.
\begin{figure}[hbt]
\hspace*{-0.2cm}\mbox{\epsfig{file=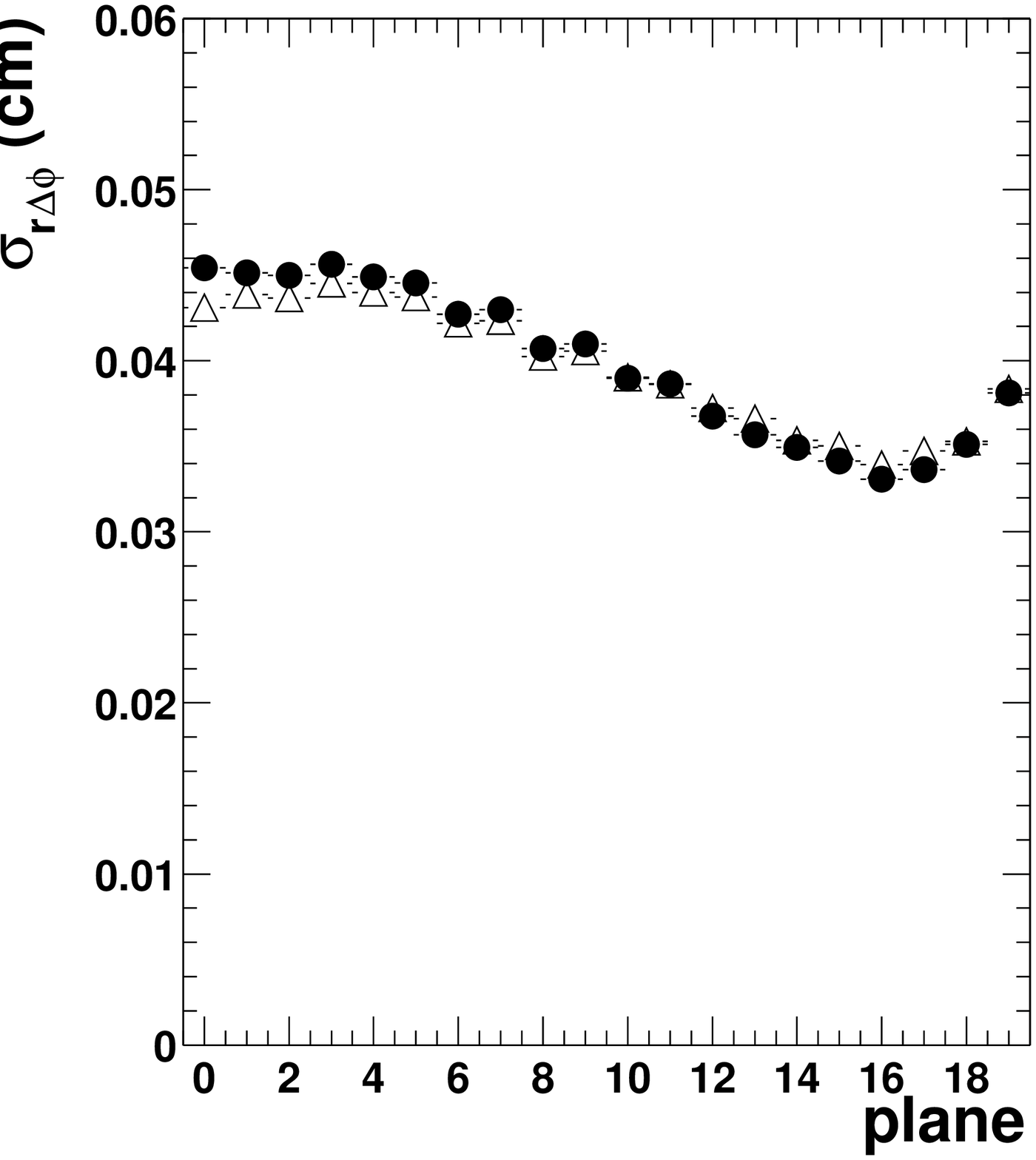,width=0.33\textwidth}}
\hspace*{0.5cm}\mbox{\epsfig{file=presthe.epsi,width=0.27\textwidth}}
\hspace*{0.6cm}\mbox{\epsfig{file=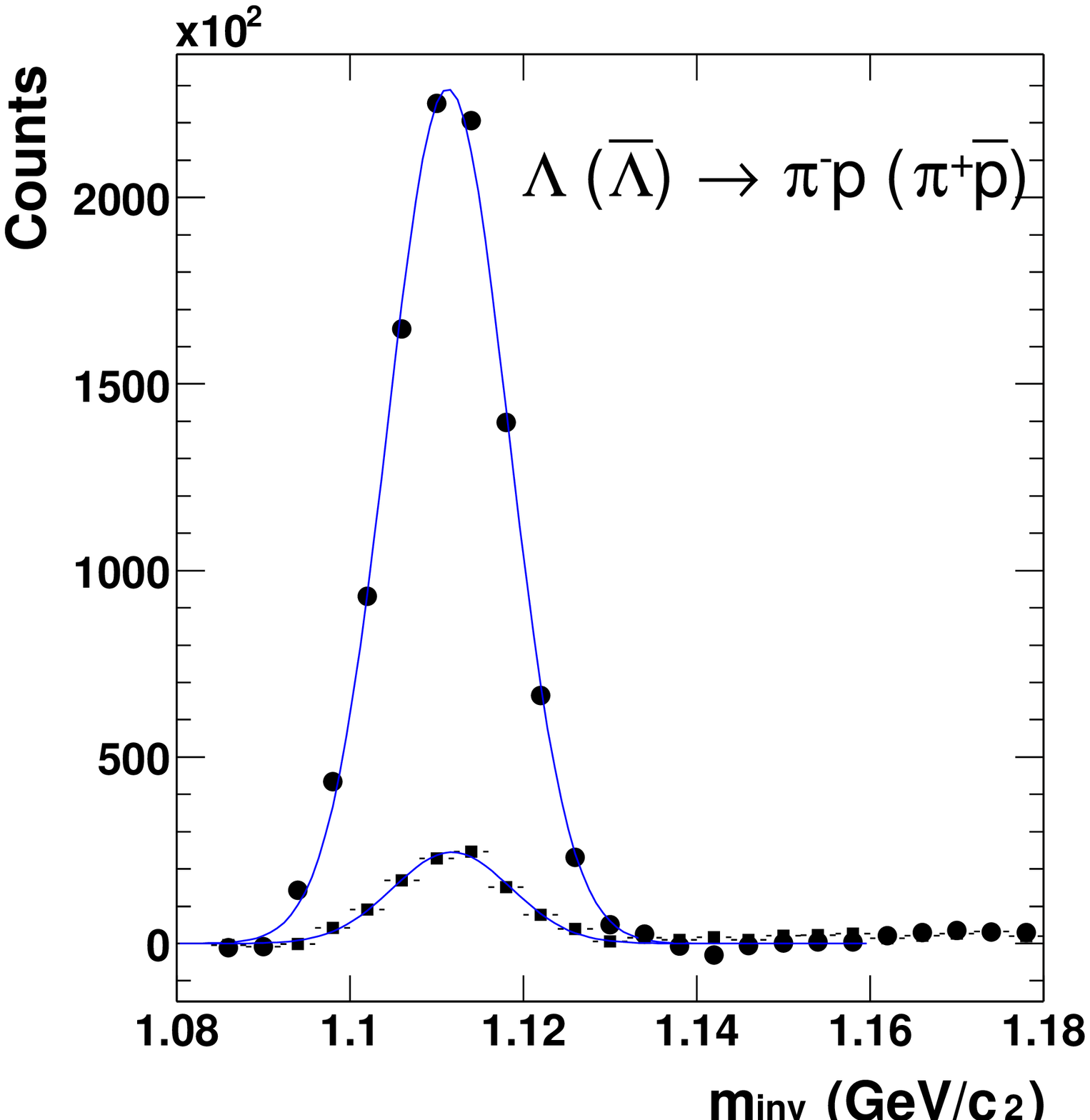,width=0.37\textwidth}}
\caption{
(Left) Single-hit position resolution in azimuthal direction (r$\Delta\phi$) as function of the
plane in the TPC ($z$) for events triggered on the 8\% (circles) and 
20\% (triangles) most central collisions obtained from hits on
tracks with a momentum $p>1$~GeV/c. (Middle) Momentum resolution from a 
simulation with single hit
position resolution as in the data for the 3 parameter fit
(triangles), for the 2 parameter fit (squares) and for the combined
fit (circles). 
(Right) Invariant mass spectra of p$\pi^-$ ($\overline{p}\pi^+$) pairs
after background subtraction for 2.0~$<y_\Lambda<$~2.4 and $p_t^\Lambda>$~1.5~GeV/c. After fitting the peaks 
of $\Lambda$ (circles) and ${\overline{\Lambda}}$ (squares) 
with a Gaussian function widths of 6.94$\pm$0.02 MeV/c$^2$ and 
6.85$\pm$0.13 MeV/c$^2$ are obtained. The measured 
${\overline{\Lambda}}$/$\Lambda$ ratio is 0.105$\pm$0.002.}
\label{fig:ka0}
\end{figure}
New hit finding and track finding algorithms were developed to better resolve
close hits and to improve tracking efficiency at very low momenta.
Tracks above 1~GeV/c momentum are found with an efficiency of
90\% due to non instrumented areas and have an average number of hits of 18.7. Down to  
a momentum of 0.3~GeV/c the efficiency is still 80\% and the average number
of hits is 17.2.
The single-hit position resolution is obtained comparing the
reconstructed hit positions with the ideal hit positions given by the
fitted trajectory of a track. The width of the residual distribution 
gives the spatial resolution. The resolution in azimuthal direction
scaled with the radius $r$ of the hit ($\sigma_{r\Delta\phi}$) 
is shown in Fig.~\ref{fig:ka0}. It deteriorates slightly with
increasing hit multiplicity in the TPC. 
The momentum resolution has also been evaluated using a microscopic drift Monte Carlo
simulation with single-point resolution as observed in the data (Fig.~\ref{fig:ka0}).
It translates into a mass resolution of about 4\% at the
$\phi$ meson mass reconstructed in the $e^+e^-$ channel.
The momentum resolution was checked in the data by looking at the width of
the reconstructed $\Lambda$ and $K^0_S$ in the invariant-mass spectrum 
(Fig.~\ref{fig:ka0}). The detailed calibration improves the mass
resolution by at least a 
factor 1.7 compared to the old calibration \cite{sqm01}.

\begin{figure}[hbt]
\mbox{\epsfig{file=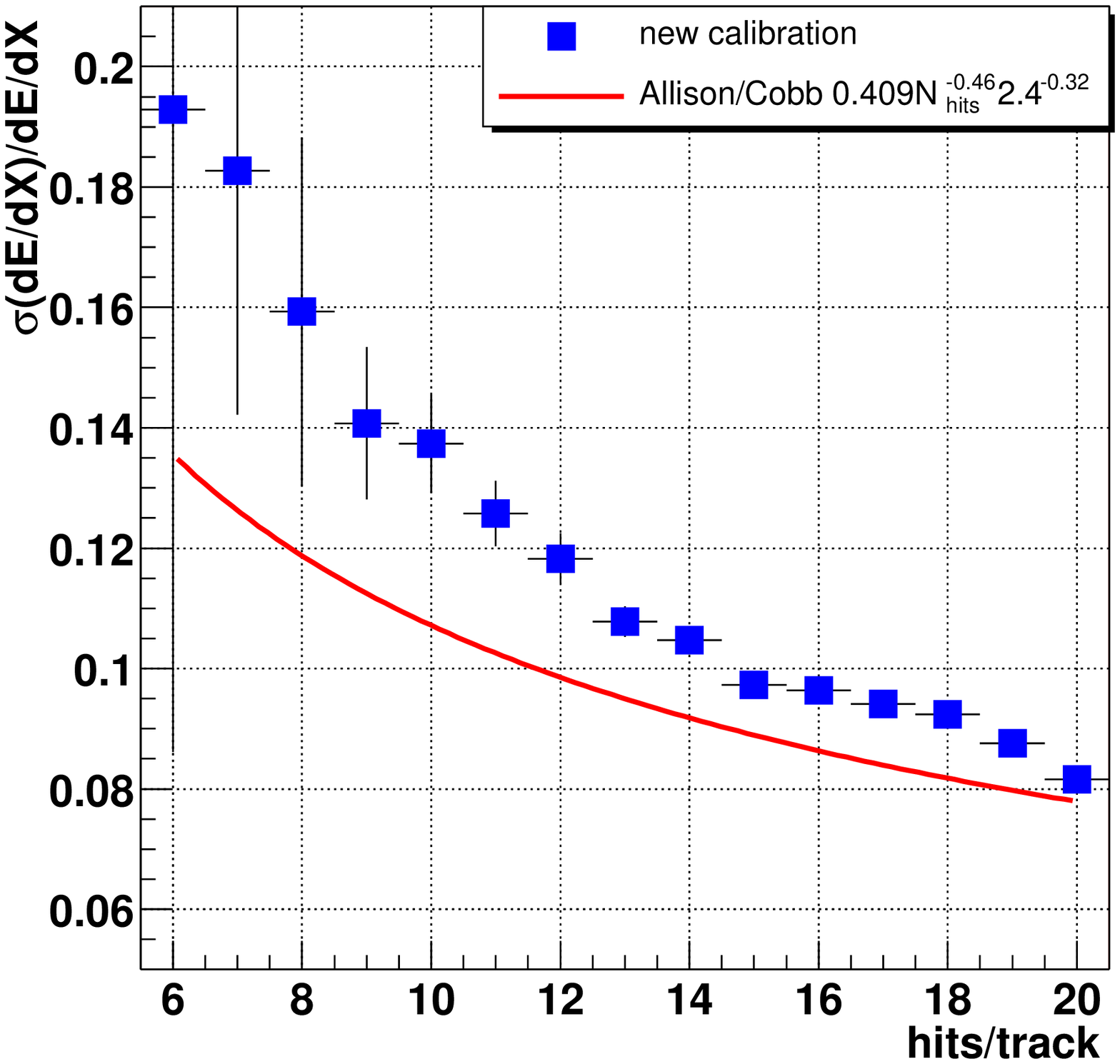,width=0.42\textwidth}}
\mbox{\epsfig{file=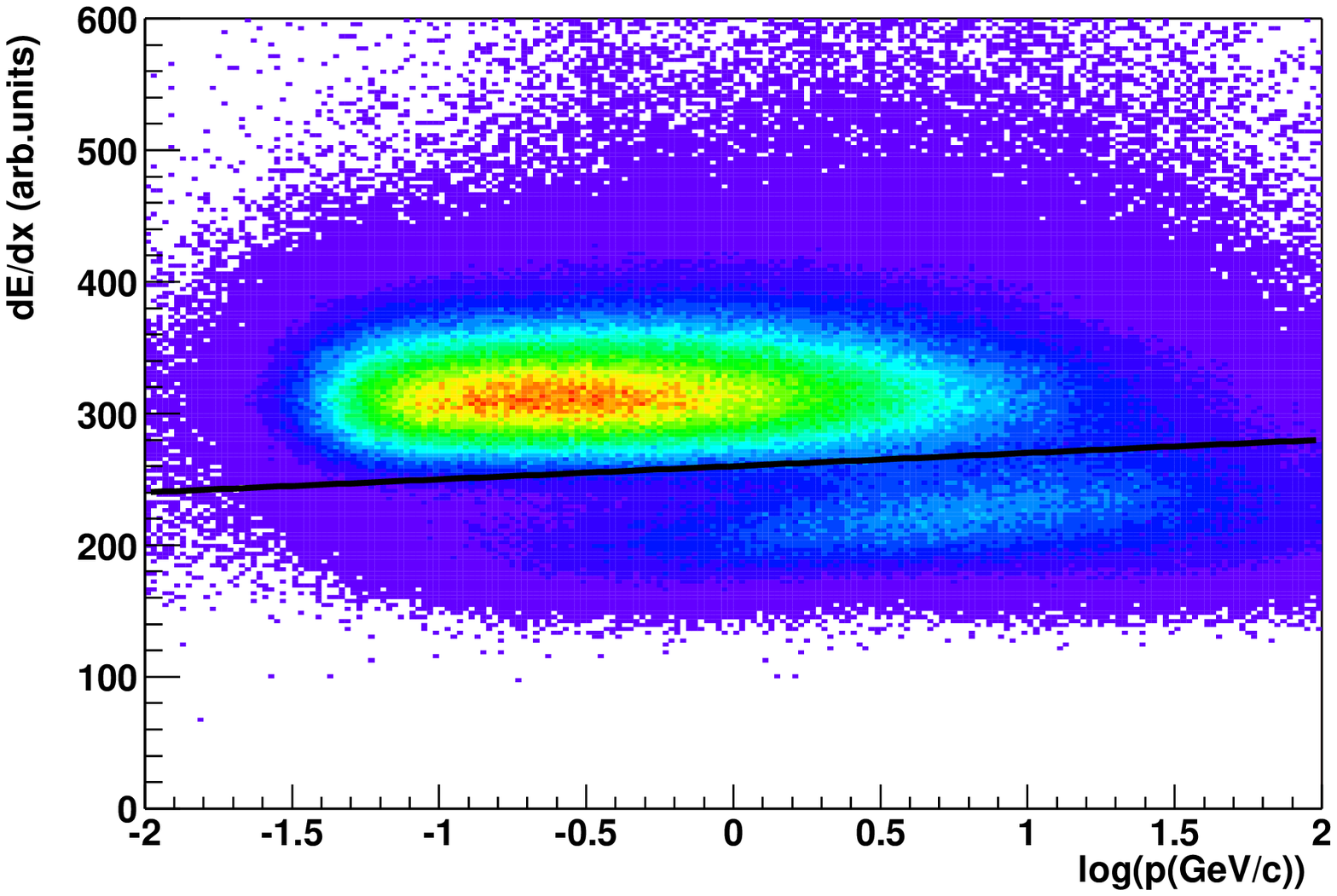,width=0.6\textwidth}}
\caption{(Left) d$E$/dx resolution of electrons as a function of number of hits in the
track after the new calibration (squares) compared to the
parameterization of Allison/Cobb (line). (Right) d$E$/dx signal in the TPC versus momentum for electrons
as selected with the RICH detector. The inclined cut against pions
(black line) show the additional rejection power from the TPC.}
\label{fig:dedx}
\end{figure}

The pad-to-pad gain calibration is obtained from the total hit amplitude in the pad
of maximum amplitude. A correction of the undershoot after each pulse obtained
from non-zero suppressed data was included \cite{lars}. An electron attachment correction
is determined by analysing the dependence of the hit amplitude on the drift
length and taking into account the different
particle abundancies for each polar angle $\theta$. 
The d$E$/dx of a track is calculated as the truncated mean of the
individual hit amplitudes. 
For the maximum number of hits on tracks the d$E$/dx resolution
approaches the parameterization by Allison and Cobb (Fig.~\ref{fig:dedx}), which implies
that most of the electrons in this analysis are measured with d$E$/dx
resolution of better than 10\%.

\section{Electron analysis}

The main difficulties of the electron analysis are the low probability of 
electromagnetic decays and the large amount of combinatorial background from $\gamma$
conversions and Dalitz decays. To select electrons among all charged hadrons
the RICH detectors and the d$E$/dx signal in the TPC are used.
The analysis chain provides track segments in the different detectors. The two
RICH detectors are used in a combined mode in this analysis. Rings with asymptotic
radius and a (double Hough) amplitude above a given threshold for the full ring
acceptance (slowly decreasing threshold towards the edges) are taken as RICH track 
segments. 
With these cuts a pion rejection of about 0.999 is achieved for an
electron efficiency of about 0.55.
Quality cuts (matching between SDD1/2 of 2 mrad and
at least 12 hits in the TPC track segments) are applied.

Global tracks are constructed by combining track segments 
in SDD, TPC and RICH with a 2$\sigma$ momentum-dependent
matching window. The electron d$E$/dx signal in the TPC is
independent of momentum (Fig.~\ref{fig:dedx}). 
A primary electron selection is done based on the d$E$/dx resolution.
Finally a more restricted cut to further suppress pions is applied (Fig.~\ref{fig:dedx}).

Very good electron identification is not enough. Electron pairs from $\gamma$
conversions and $\pi^0$ Dalitz decays need to be identified by
topology and removed from the sample to reduce combinatorial background. They are 
characterized by their small opening angle and low momentum. 
As the detectors have a
finite two-track resolution the most effective way of rejecting
conversions in the target and close Dalitz pairs is
by rejecting tracks with at least twice the minimum ionizing energy
loss signal in both SDD's where all
hit amplitudes have been re-summed within 7 mrad around the each SDD track
segment. Low-amplitude tails in the SDD d$E$/dx distribution are also removed. 
Late conversions (mostly in SDD1) are removed by an
upper cut in each RICH ring amplitude and by a cut in the distance to
another TPC track of opposite sign and electron d$E$/dx. 
\begin{wrapfigure}{r}{6.5cm}
\hspace*{-0.4cm}\mbox{\epsfig{file=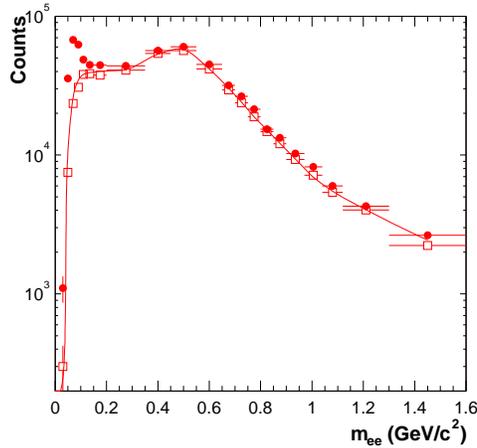,width=0.55\textwidth}}
\label{fig:fullrejectmix}
\caption{ Invariant-mass distribution of unlike-sign pairs (full symbols), 
like-sign pairs (open symbols), and mixed-event background (line) normalized
to like-sign pairs background after full rejection.}
\end{wrapfigure}
To reduce $\pi^0$ Dalitz decay contributions only electron tracks 
with no opposite charge electron track within 35 mrad are taken for
analysis. Finally, only electron tracks in the geometrical 
acceptance 0.14~rad~$< \theta< $~0.243 rad and with
a $p_t>$~0.2~GeV/c are selected. 

The invariant-mass distributions of unlike-sign pairs and of like-sign pairs that is a
measure of the combinatorial background after full rejection are shown in
Fig.~\ref{fig:fullrejectmix}. In order to reduce the statistical
errors an unlike-sign combinatorial background using the mixed-event technique
has also been evaluated.
For masses below 0.2~GeV/c$^2$, the mixed-event background deviates
from the like-sign background due to ring reconstruction effects; the mixed-event background is
normalized to the like-sign pair background in the mass region 0.2~GeV/c$^2<$m$_{e^+e^-}<$0.6~GeV/c$^2$.
The physics signal is obtained by subtracting the like-sign pairs or
the mixed event background from the unlike-sign pairs 
as $N_{e^+e^-}-(N_{e^+e^+}+N_{e^-e^-})$. This analysis is based on
about $18\cdot 10^6$ events corresponding aproximately to 80\% of the events
where both RICH detectors were operational.

\section{Low-mass e$^+$e$^-$ results}
\begin{figure}[ht]
\hspace*{-0.4cm}\mbox{\epsfig{file=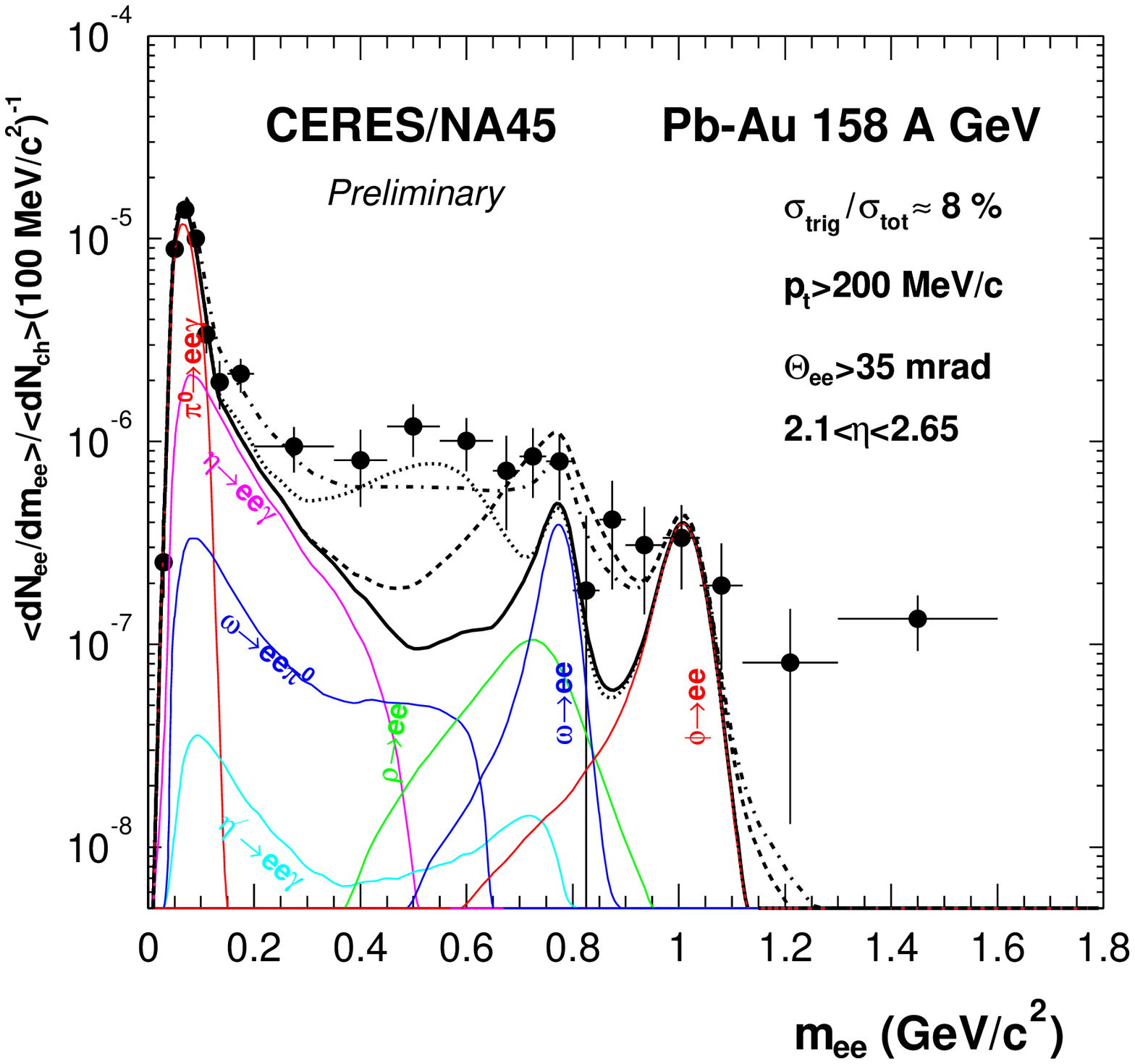,width=0.55\textwidth}}
\hspace*{-0.4cm}\mbox{\epsfig{file=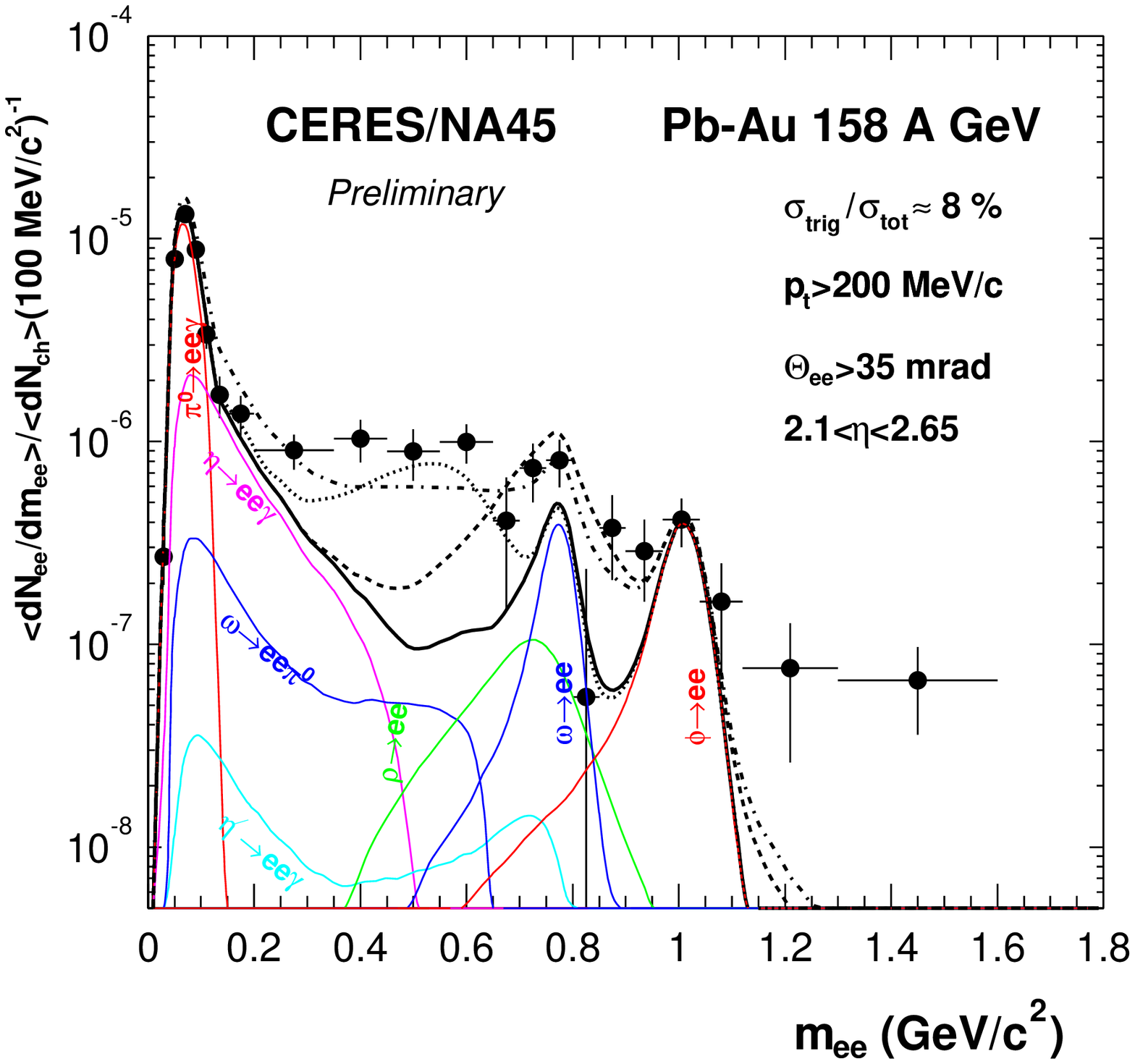,width=0.55\textwidth}}
\caption{Invariant-mass spectrum of $e^+e^-$-pairs
normalized to the expectation from the hadron decay 
cocktail (thick solid line) in the $\pi^0$ Dalitz peak using the
like-sign background (left) or the mixed-event background (right). The
expectations from model calculations assuming
the vacuum $\rho$ spectral function (dashed), a dropping mass (dotted) or a in medium spread $\rho$ width (dashed-dotted) are also shown.
}
\label{fig:fullreject}
\end{figure}

The preliminary e$^+$e$^-$ invariant-mass spectrum normalized to the
$\pi^0$ Dalitz peak of the hadronic decay cocktail is shown in 
Fig.~\ref{fig:fullreject}. The hadron decay cocktail
\cite{cocktail} has been folded with the experimental momentum resolution 
and energy loss due to bremsstrahlung. 
Acceptance, opening-angle, and transverse-momentum cuts are applied.
An excess of pairs for m$_{e^+e^-}>$ 0.2~GeV/$c^2$ is visible.
Using the like-sign background, the number of pairs in the Dalitz region (m$_{e^+e^-}<$~0.2~GeV/c$^2$) 
is 2826~$\pm$~114 with a signal to background ratio (S/B) of
1/1.78. Using the mixed-event background, the number of open 
pairs (m$_{e^+e^-}>$~0.2~GeV/c$^2$) is 2023~$\pm$~182 with a S/B of 1/14.6.
The enhancement factor for 0.2~GeV/c$^2$$<$ m$_{e^+e^-}$$<$1.1~GeV/c$^2$ compared to the 
hadron decay cocktail is 3.1 $\pm$ 0.3 (stat). 
Further optimization of cuts based on Monte Carlo
simulations is in progress and is expected to improve the statistical errors.

The experimental results are compared to theoretical
models based on hadronic decays and $\pi^+\pi^-$ annihilation. 
The $\rho$-propagator is treated \cite{rapp} in 3 ways: vacuum $\rho$, modifications 
following Brown-Rho scaling \cite{theo2}, and modifications via $\rho$-hadron
interactions \cite{theo1}. The data clearly disfavour the vacuum $\rho$.
In the region between the $\omega$ and the $\phi$ the data seem to 
favour the many-body approach over Brown-Rho scaling.
We have to await the final results of the analysis before
a stronger statement can be made.

\section{The $\phi$ meson puzzle}

One of the open issues at the CERN SPS is the $\phi$ puzzle. Two
experiments, one measuring the $\phi$ meson in the $\mu^+$ $\mu^-$
decay channel (NA50), and the other (NA49), measuring it in the $K^+K^-$ channel,
obtain yields that differ by factors between 2 and 4 in the common 
$m_t$ range \cite{puzzle}. Further, the $m_t$ spectra exhibit a different inverse 
slope parameter \cite{puzzle}, 305$\pm$15 MeV in NA49 and 218 $\pm$ 6 MeV in NA50,
fitted in their $m_t$ acceptance regions. 

\begin{figure}[htb]
\hspace*{-0.3cm}\mbox{\epsfig{file=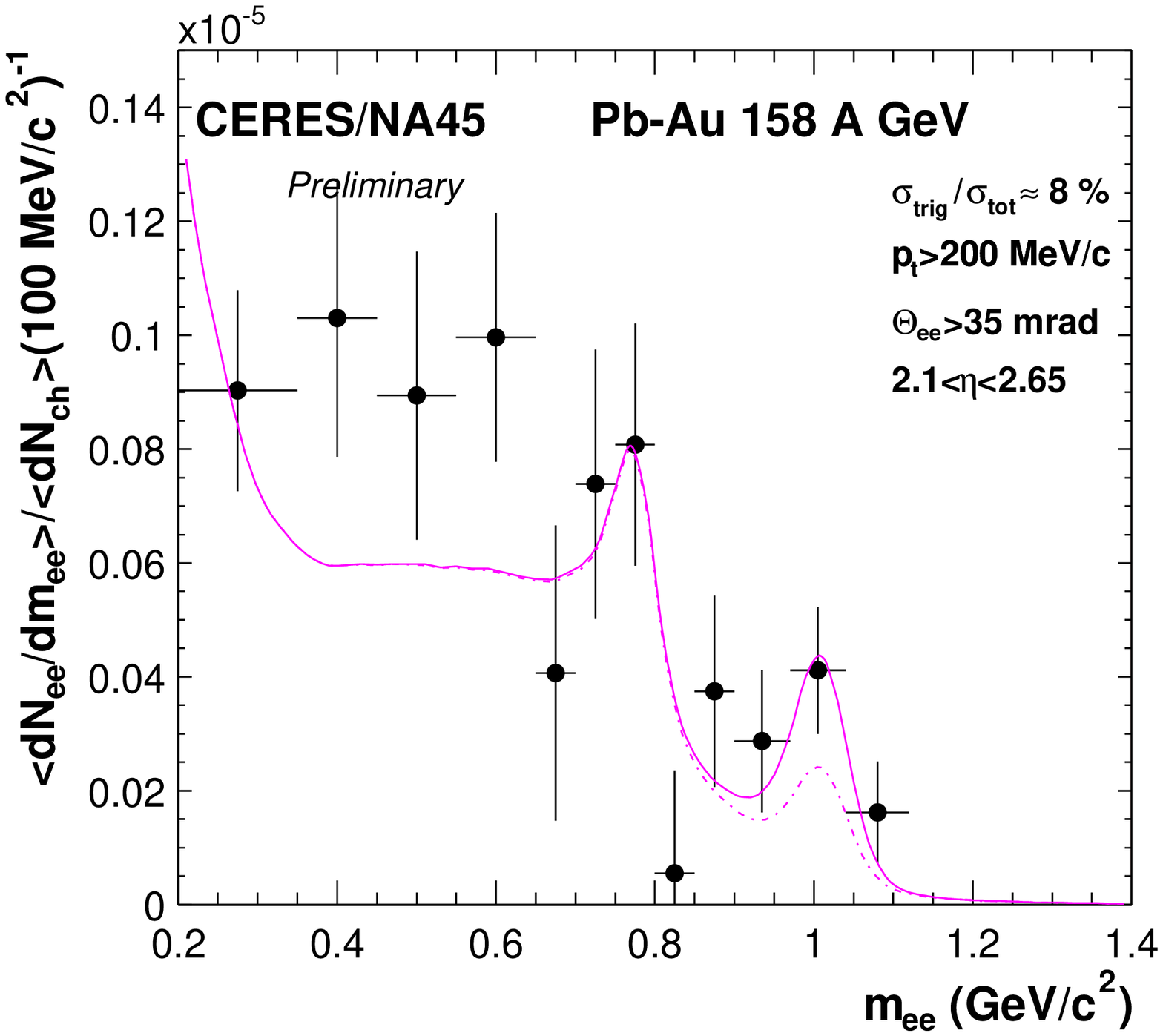,width=0.505\textwidth}}
\mbox{\epsfig{file=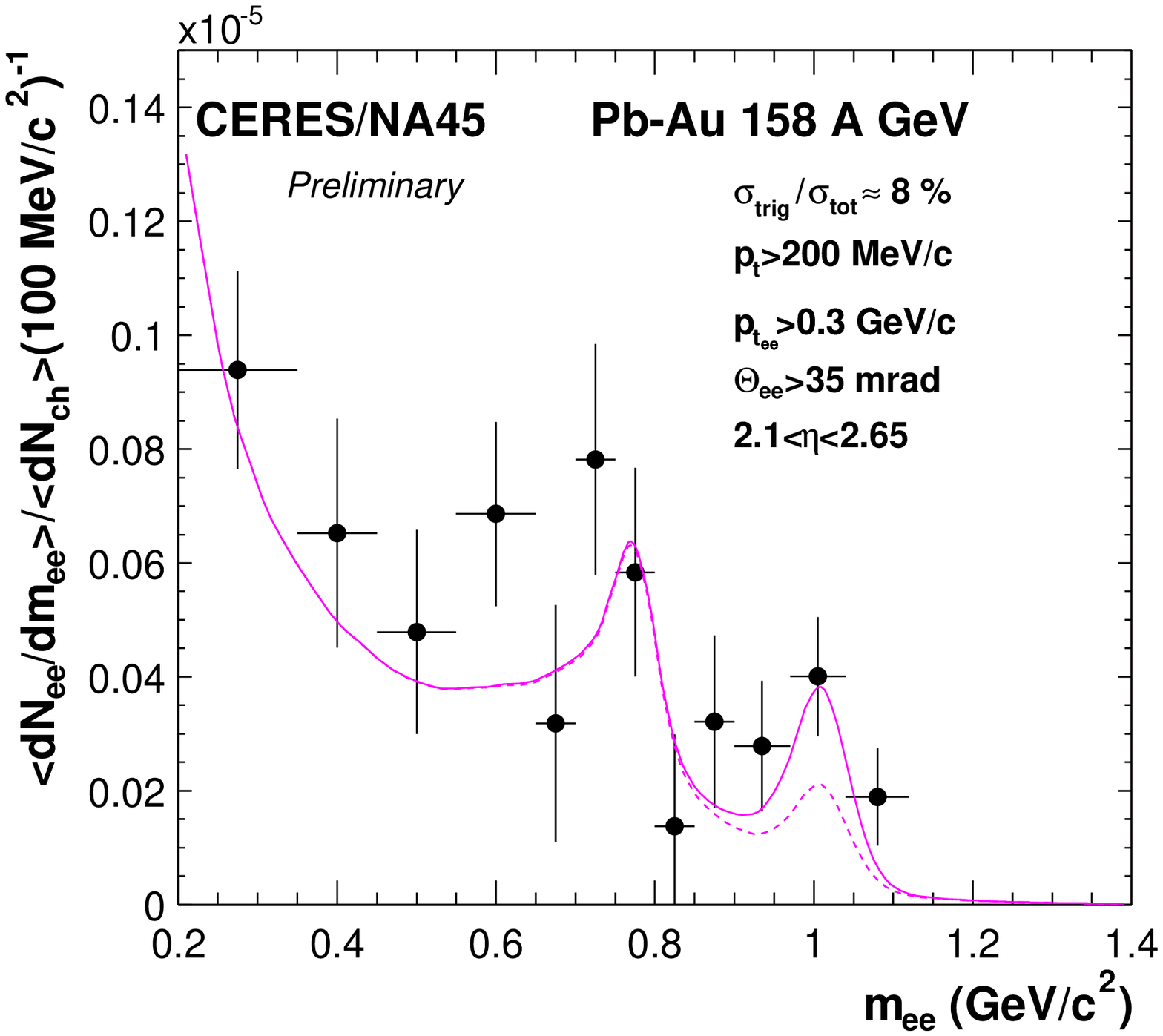,width=0.505\textwidth}}
\vspace*{-0.3cm}
\caption{Invariant-mass spectrum of $e^+e^-$-pairs
together with the
expectations from model calculations assuming in-medium spread $\rho$
width plus the thermal $\phi$ yield (thin solid) or the thermal yield
reduced to 50\% (dashed) are shown. The left plot corresponds to the
standard set of cuts and the right one has an additional electron pair $p_{t}$ cut
of 0.3 GeV/c.} 
\label{fig:phi_ele}
\end{figure}
We have compared our preliminary mass spectrum with a calculated
spectrum based on a $\phi$-yield as predicted in
a thermal model \cite{thermal} which also describes the
NA50 yield (left of Fig.~\ref{fig:phi_ele}). The lines also
contain the $\rho$-contribution. Within a window of $\pm 1\sigma$ around the
$\phi$ peak the ratio of data to simulation is 1.1~$\pm$~0.3. With an
additional electron pair $p_{t}$ cut of 300 MeV/c (right of Fig.~\ref{fig:phi_ele}) the ratio is 
1.2~$\pm$~0.3. Obviously this is in very good agreement with our data.
Reducing the $\phi$-yield to 50\% of the above value (dashed 
lines of Fig.~\ref{fig:phi_ele}) leads to a ratio of 1.9~$\pm$~0.5 and
2.1~$\pm$~0.6. The reduced yield that would agree with the NA49 observation for
$\phi\rightarrow K^+K^-$ seems in disagreement with our preliminary data. 
The evaluation of absolute efficiencies and systematic errors are
needed to draw a definitive conclusion.

The upgraded spectrometer offers the unique possibility to study 
simultaneously the $\phi$ meson in the $K^+K^-$ decay channel.
To study this channel, we assign to all charged particles the kaon mass 
(no particle identification). TPC tracks in the geometrical acceptance 
0.14~rad $<\theta<$~0.24~rad with at least 12 hits and with a
transverse momentum $p_t > 0.25$~GeV/c that are matched to a SDD
track within 2.5$\sigma$ momentum-dependent matching window are used.
The combinatorial background is calculated using the mixed-event
technique. An opening-angle cut of 0.015~rad~$<\theta_{12}<$~0.17~rad
is applied to improve the signal-to-background ratio. 
A cut $q_t<0.19-0.4\cdot\alpha^4-0.4\cdot\alpha^2$, where $\alpha$ is
the Podolanski-Armenteros parameter
$\alpha=(p_L^+-p_L^-)/(p_L^++p_L^-)$  and $p_L^{\pm}$ ($q_t$) is
the decay track momentum component parallel (transverse) to the
$\phi$ momentum in the laboratory frame, is also applied
to avoid contamination 
from other particles species. The signal is studied 
for $p_t^\phi>1$ GeV/c and rapidity 2.0 $<y^\phi<$ 2.4. 
Acceptance and pair reconstruction
efficiency corrections are obtained from GEANT simulations where $\phi$ decays
with a distribution in transverse momentum 
$dN/dp_t = C\cdot p_t \cdot exp({-m_t/T})$ and a
Gaussian rapidity distribution with a standard deviation of 1.2 
are embedded into the background of real events.
\begin{figure}[htb]
\mbox{\epsfig{file=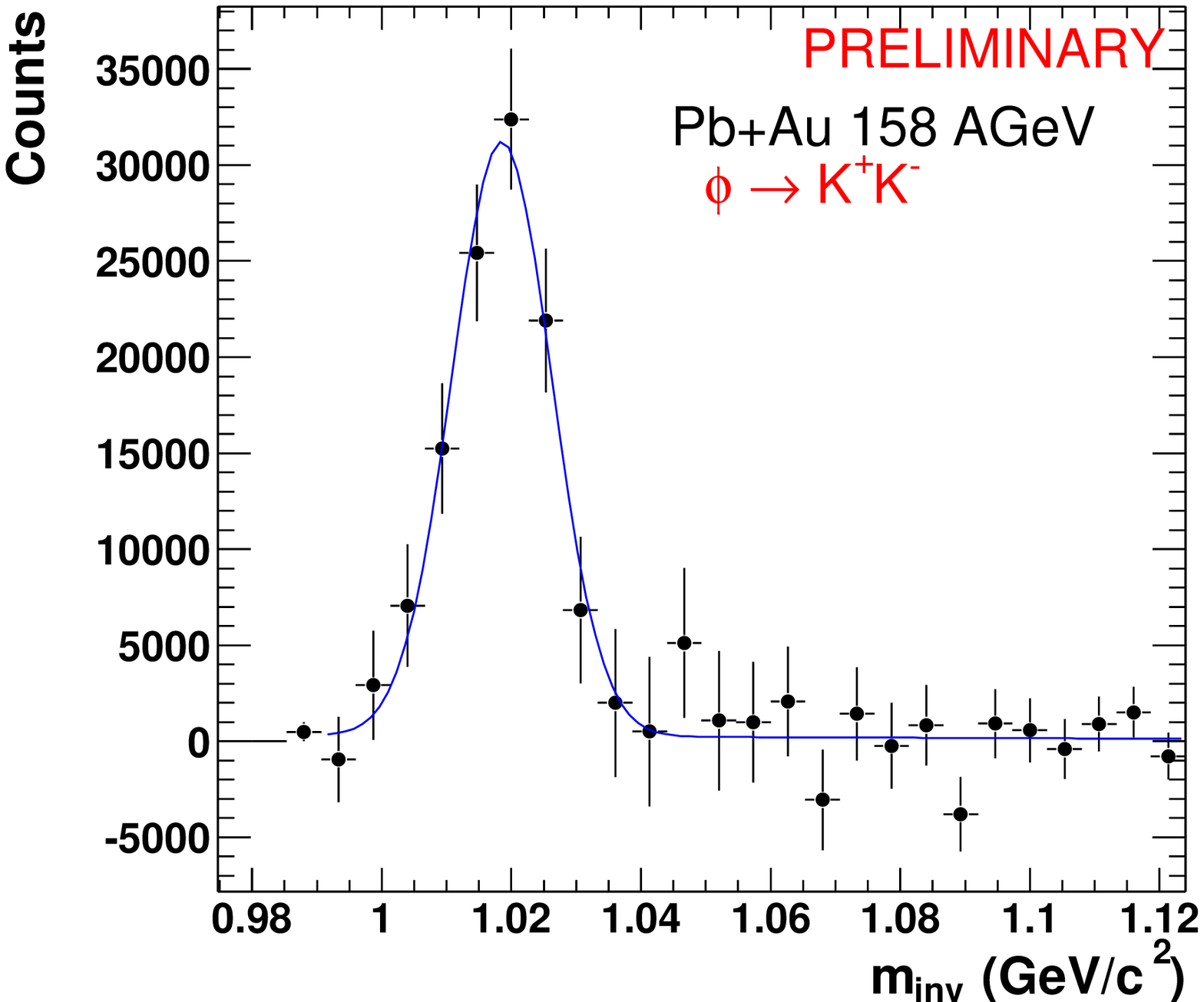,width=0.49\textwidth}}
\mbox{\epsfig{file=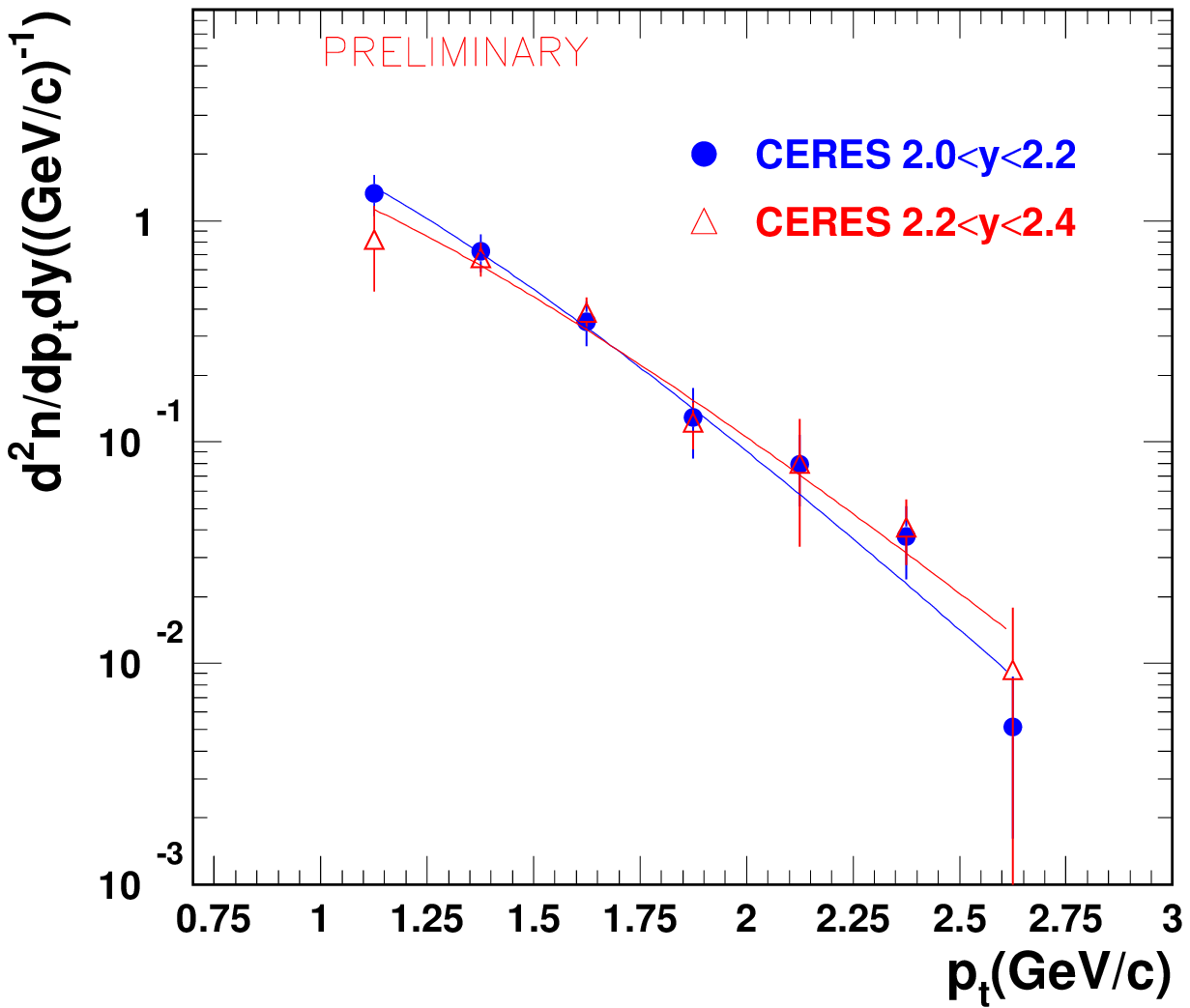,width=0.51\textwidth}}
\caption{Preliminary invariant-mass spectrum of $K^+K^-$ pairs after background
subtraction for 2.2~$<y_\phi<$~2.4 and 1.5~GeV/c~$<p_t^\phi<$~1.75~GeV/c . A peak corresponding to the $\phi$ meson is observed
(left). Transverse momentum spectrum of the $\phi$ meson corrected for
acceptance and efficiency. Also shown are fits with a mentioned
distribution. For fits see text.}
\label{fig:phi_ka}
\end{figure}
The preliminary invariant mass
spectrum after background subtraction and the transverse momentum spectra corrected
for efficiency and acceptance are presented in Fig.~\ref{fig:phi_ka}.
The values of the inverse slope parameter obtained after fitting with
the function given above are T=218$\pm$12 MeV and T=246$\pm$22 MeV for the
rapidity intervals 2.0$<$y$<$2.2 and 2.2$<$y$<$2.4, respectively.
Evaluation of systematic errors and of the exact centrality necessary to
compare to other experiments are in progress.


\section*{References}
\begin{thebibliography}{99}

\bibitem{exp} G.~Agakichiev {\it et al.} CERES Collaboration {\it Phys. Rev. Lett.} 
75 (1995) 1272. 


\bibitem{nantes}G.~Agakichiev {\it et al.} CERES Collaboration {\it
Phys. Lett. B} 422 (1998) 405; B.~Lenkeit for the  CERES Collaboration
{\it Nucl. Phys.} A 661 (1999) 23c. J.P.~Wessels for the  CERES
Collaboration {\it Nucl. Phys.} A 715 (2003) 607c.

\bibitem{40gev}D.~Adamova et al. {\it  Phys. Rev. Lett.} CERES
Collaboration 91 (2003) 042301.

\bibitem{pp} G.~Agakichiev {\it et al.} CERES Collaboration {\it Eur. Phys. Jour. C} 4 (1998) 231. 

\bibitem{up2} Addendum to proposal SPSLC/P280: CERN/SPSLC 96-35/P280 Add.1.

\bibitem{up1} Technical Note on the NA45/CERES upgrade. CERN/SPSLC 96-50 (1996).

\bibitem{qm99} A. Mar\'{\i}n for the CERES Collaboration {\it Nucl. Phys.} A 661 (1999) 673c.

\bibitem{qm01} H. Appelsh\"auser for the CERES Collaboration {\it Nucl. Phys. A}698 (2002) 253c.

\bibitem{hbt} D.~Adamova {\it et al.} CERES Collaboration {\it  Nucl. Phys.} A714 (2003) 124.
\bibitem{hbt1} D.~Adamova {\it et al.} CERES Collaboration {\it Phys. Rev. Lett.} 90 (2003) 022301.
\bibitem{fluc} D.~Adamova {\it et al.} CERES Collaboration {\it Nucl. Phys.A} 727 (2003) 97.
\bibitem{fluc1} H.~Sako CERES Collaboration, these proceedings.

\bibitem{gar} R. Veenhof, {\it Nucl. Instr. and Meth.} 419 (1998) 726;
{\it http://consult.cern.ch/writeup/garfield/}. S.F.~Biagi, {\it Nucl. Instr. and Meth.} 421 (1999) 234.
\bibitem{sqm01} W. Schmitz for the CERES Collaboration {\it Jour. Phys. G} 28 (2002) 1861.
\bibitem{lars} L. Dietrich, Diploma Thesis, University of Heidelberg (2001).

\bibitem{allison} W.W.M.~Allison and J.H.~Cobb Ann. Rev. Nucl. Part. Sci. 30 (1980) 253. 

\bibitem{cocktail} H.~Sako for the CERES Collaboration. Technical Report 03-25 (2000).

\bibitem{rapp} R. Rapp, private comunication.

\bibitem{theo2} G.E.~Brown and M.~Rho. Phys. Rep. 363 (2002) 85.

\bibitem{theo1} R.~Rapp and J. Wambach, Adv. Nucl. Phys. 25 (2000) 1; private comunication.

\bibitem{puzzle} D.~R\"ohrich {\it J. Phys. G } 27 (2001) 355.
\bibitem{thermal}P.~Braun-Munzinger, I.~Heppe and J.~Stachel {\it Phys. Lett.B} 465 (1999) 15.



\end{thebibliography}
\end{document}